\documentclass{gretsi}

  \usepackage{times}            

\usepackage[ansinew]{inputenc}
\usepackage[english,francais]{babel}
\usepackage{times}
\usepackage{amsfonts}
\usepackage{amstext}
\usepackage{amssymb}
\usepackage{graphics}
\usepackage{epsfig,amsmath}

\titre{Méthode du point proximal: principe et applications aux
algorithmes itératifs\\}

\auteur{\coord{Ziad}{Naja}{1},
        \coord{Florence}{Alberge}{1},
    \coord{Pierre}{Duhamel}{2}}

\adresse{\affil{1}{Univ Paris-Sud 11
         }
         \affil{2}{CNRS \\
         Laboratoire des signaux et syst\`{e}mes (L2S)\\
         Supelec, 3 rue Joliot-Curie 91192 Gif-sur-Yvette cedex (France)}}

\email{Ziad.Naja@lss.supelec.fr, Florence.Alberge@lss.supelec.fr,
Pierre.Duhamel@lss.supelec.fr}

\resumefrancais{Cet article est basé sur l'algorithme du point
proximal. Nous étudions deux algorithmes itératifs: l'algorithme de
Blahut-Arimoto communément utilisé pour le calcul de la capacité des
canaux discrets sans mémoire puis le décodage itératif pour les
modulations codées à bits entrelacés. Dans les deux cas, il s'agit
d'algorithmes itératifs pour lesquels les méthodes de type point
proximal conduisent à une nouvelle interprétation et ouvrent la voie
à des améliorations en terme de vitesse de convergence notamment.}

\resumeanglais{This paper recalls the proximal point method. We
study two iterative algorithms: the Blahut-Arimoto algorithm for
computing the capacity of arbitrary discrete memoryless channels, as
an example of an iterative algorithm working with probability
density estimates and the iterative decoding of the Bit Interleaved
Coded Modulation (BICM-ID). For these iterative algorithms, we apply
the proximal point method which allows new interpretations with
improved convergence rate.}

\begin{document}
\maketitle \vspace{-2cm}
\section{Introduction}
Cet article s'intéresse à deux algorithmes itératifs classiques:
l'algorithme de Blahut-Arimoto \cite{Arimoto, Blahut} pour le calcul
de la capacité d'un canal discret sans mémoire et le décodage
itératif des modulations codées à bits entrelacés (BICM-ID)
\cite{Caire}. Bien que ces méthodes soient radicalement différentes
à la fois par l'application visée et aussi par le processus itératif
mis en jeu, elles ont pour point commun de présenter des connections
avec une méthode d'optimisation bien connue, la méthode du point
proximal \cite{Vige}.\\En 1972, R. Blahut et S. Arimoto
\cite{Arimoto,Blahut} ont montré comment calculer numériquement la
capacité des canaux sans mémoire avec des entrées et des sorties à
alphabets finis. Depuis, plusieurs extensions ont été proposées
citons notemment \cite{Dupuis04} qui a étendu l'algorithme de
Blahut-Arimoto aux canaux avec mémoire et entrées à alphabets finis
et \cite{Dauwels} qui a considéré des
canaux sans mémoire avec des entrées et/ou des sorties continues.\\
En parallèle, d'autres travaux se sont concentrés sur
l'interprétation géométrique de l'algorithme de Blahut-Arimoto
\cite{Csiszar_proj}. En se basant sur cette dernière approche, Matz
\cite{Matz04} a proposé une version modifiée de cet algorithme qui
converge plus vite que l'algorithme standard.\\L'algorithme proposé
par Matz est basé sur une approximation d'un algorithme de point
proximal. Nous proposons donc dans ce qui suit une vrai
reformulation point proximal avec une vitesse de convergence plus
grande comparée à celle de l'algorithme classique de Blahut-Arimoto
ainsi qu'à celle de l'approche dans \cite{Matz04}.\\D'autre part,
les modulations codées à bits entrelacés (BICM) ont été d'abord
proposés par Zehavi \cite{Zehavi} pour améliorer la performance des
modulations codées en treillis dans le cas des canaux de Rayleigh à
évanouissement. Le décodage itératif \cite{Li2} utilisé pour les
BICM a une structure similaire à celle d'un turbo décodeur série.
Bien que très performant, le décodage itératif n'a pas été à l'origine introduit comme solution d'un problème d'optimisation, ce qui rend difficile l'analyse de sa convergence.\\
Cet article va donc mettre en évidence le lien existant entre ces
deux algorithmes itératifs et montrer comment cela conduit à des
améliorations substantielles tout en révélant le lien existant entre
le décodage itératif et les techniques classiques d'optimisation.
\vspace{-0.4cm}
\section{Algorithme du point proximal}
L'algorithme du point proximal, dans sa version d'origine, est
caractérisé par le processus itératif \cite{Hero}:
\begin{equation}
\label{proxi2}\theta^{(k+1)}=\arg\max_{\theta}\{\xi(\theta)-\beta_k
\|\theta-\theta^{(k)}\|^2\}
\end{equation}
dans lequel $\xi(\theta)$ est la fonction de coût qui croît au fil
des itérations et $\|\theta-\theta^{(k)}\|^2$ est un terme de
pénalité qui assure que la nouvelle valeur du paramètre reste dans
le voisinage de la valeur obtenue à l'itération précédente.
$\{\beta_k\}_{k\geq 0}$ est une séquence de paramètres positifs.
lorsque la séquence ${\beta_k}$ converge vers zéro à l'infini, alors
la méthode présente une convergence super-linéaire
\cite{Rockafellar}. L'algorithme du point proximal peut être
généralisé selon:
\begin{center}
$\theta^{(k+1)}=\displaystyle{\arg\max_\theta}\{\xi(\theta)-\beta_k
f(\theta,\theta^{(k)})\}$
\end{center}
où $f(\theta,\theta^{(k)})$ est toujours non négative et
$f(\theta,\theta^{(k)})=0$ si et seulement si $\theta=\theta^{(k)}$.
Dans la suite, nous utiliserons cette formulation en considérant
pour $f$ soit la divergence de Kullback soit la divergence de
Fermi-Dirac. Nous rappelons maintenant leurs définitions.\\La
distance de Kullback-Leibler (KLD) est définie pour deux
distributions de probabilité $ \textit{p}=\{p(x),x \in
\textit{\textbf{X}}\}$ et $\textit{q}=\{q(x),x \in
\textit{\textbf{X}}\}$ d'une variable aléatoire discrète \textbf{X}
prenant ses valeurs \textbf{x} dans un ensemble discret
\textit{\textbf{X}} par:
\begin{displaymath}
\textit{D}(\textit{p}||\textit{q})=\sum_{\textbf{x}\in\textit{\textbf{X}}}{p(\textbf{x})\log\frac{p(\textbf{x})}{q(\textbf{x})}}
\end{displaymath}
La distance de Kullback (appelée aussi entropie relative) a deux
propriétés importantes: $D(p||q)$ est toujours non-négative, et
$D(p||q)$ est nulle si et seulement si $p=q$. Cependant, ce n'est
pas une "vraie" distance puisqu'elle n'est pas symétrique\\
($D(\textit{p}||\textit{q})\neq D(\textit{q}||\textit{p})$) et ne
satisfait pas en général l'inégalité triangulaire.\\ La divergence
de Fermi-Dirac est la divergence de Kullback-Leibler appliquée à des
probabilités sur des évènements
n'ayant que deux issues, elle est définie pour deux distributions de probabilité $r_i=P_R(x_i=1)$ et $s_i=P_S(x_i=1)$ définies dans l'ensemble $\textbf{X}=(x_1,\ldots, x_n)$ avec $x_i\in\{0,1\}$ de la manière suivante:\\
$D_{FD}({\bf r},{\bf s})= \sum_{i=1}^{n} r_i
\log\left(\frac{r_i}{s_i}\right) + \sum_{i=1}^{n}
(1-r_i)\log\left(\frac{1-r_i}{1-s_i}\right)$\\La divergence de
Fermi-Dirac présente les deux mêmes propriétés que la distance de
Kullback: $D_{FD}({\bf r},{\bf s})$ est toujours non négative et
$D_{FD}({\bf r},{\bf s})=0$ si et seulement si ${\bf r}={\bf s}$. La
divergence de Fermi-Dirac n'est pas symétrique. \vspace*{-0.4cm}
\section{Méthode de point proximal pour les algorithmes itératifs}
\subsection{Algorithme de Blahut-Arimoto \cite{Arimoto} et\\
interprétation point proximal}
\begin{figure}[!h]
\centerline{\epsfxsize=5cm\epsfysize=1cm\epsfbox{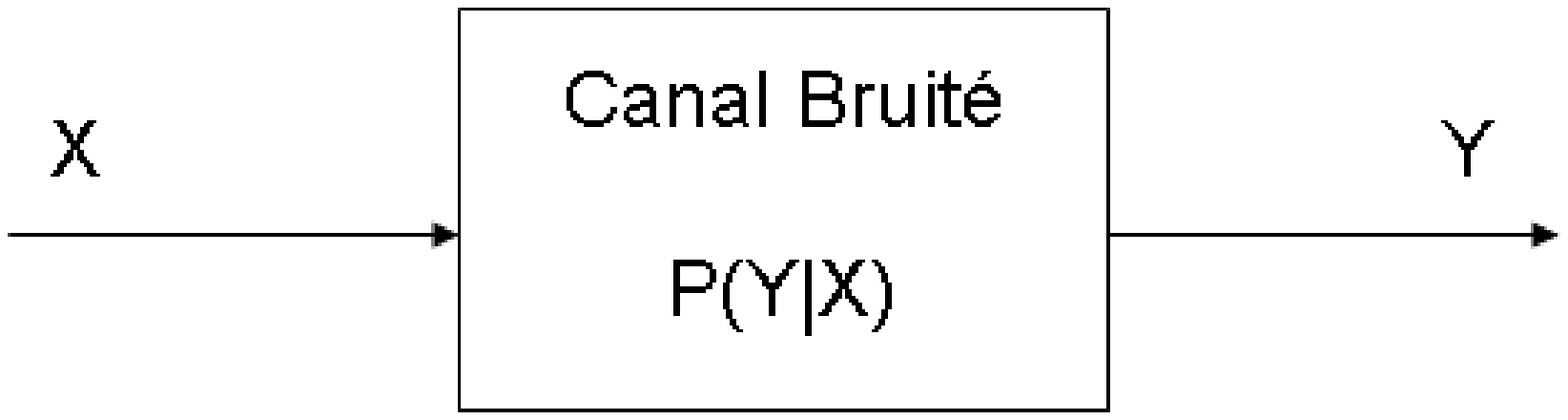}}\vspace*{-0.5cm}\caption{\footnotesize
\label{result1} Canal.}
\end{figure}
Considérons un canal discret sans mémoire avec pour entrée X prenant
ses valeurs dans l'ensemble $\{x_0,\ldots,x_M\}$ et en sortie Y
prenant ses valeurs dans l'ensemble $\{y_0,\ldots,y_N\}$. Ce canal
est défini par sa matrice de transition Q telle que $[Q]_{ij}$ =
$Q_{i|j}=Pr(Y=y_i|X=x_j)$.\\ Nous définissons aussi $p_j=Pr(X=x_j)$
et $q_i=Pr(Y=y_i)$. L'information mutuelle est donnée par:
$\textit{I(X,Y)}=\textit{I(p,Q)}=\sum_{j=0}^{M}\sum_{i=0}^{N}{p_j
Q_{i|j} \log\frac{Q_{i|j}}{q_i}}=\sum_{j=0}^{M}{p_j D(Q_j||q)}$
et la capacité du canal par:
$$C=\max_{p}{I(p,Q)}$$
En résolvant ce problème de maximisation et en prenant en compte la
condition de normalisation, nous obtenons le processus itératif:
\begin{equation}
\label{BA} p^{(k+1)}{(x)}=\frac{p^{(k)}{(x)}
\exp(D_x^k)}{\sum_{x}^M{p^{(k)}{(x)} \exp(D_x^k)}}
\end{equation}
avec $D_x^k=D(p(Y=y|X=x)||p(Y=y^{(k)}))$. C'est l'algorithme de
Blahut-Arimoto. On peut montrer sans difficulté que cet algorithme
est équivalent à:
\begin{equation}
\label{proxi1}p^{(k+1)}{(x)}=\arg\max_p\{I^{(k)}{(p(x))}-D(p(x)||p^{(k)}{(x)})\}
\end{equation}
où $I^{(k)}{(p(x))}=\mathbb{E}_{p(x)}\{D_x^k\}$. Cet algorithme
n'est pas un algorithme du point proximal puisque la fonction de
co\^{u}t $I^{(k)}{(p(x))}$ dépend des itérations. Il est toutefois
possible d'exprimer l'information mutuelle comme suit:
\begin{equation}
\label{proxi3}I(p(x))=I^{(k)}{(p(x))}-D(q(y)||q^{(k)}{(y)})
\end{equation}
En introduisant (\ref{proxi3}) dans (\ref{proxi1}), nous obtenons:
\begin{small}
$$p^{(k+1)}{(x)}=\arg\max_p\{I(p(x))-(
D(p(x)||p^{(k)}{(x)})-D(q(y)||q^{(k)}{(y)}))\}$$
\end{small}
D'après l'inégalité de Jensen, nous pouvons montrer que le terme de
pénalité
\begin{eqnarray}
D(p(x)||p^{(k)}{(x)})-D(q(y)||q^{(k)}{(y)}) & = &
\nonumber\\\mathbb{E}_{p(x,y)}{[\log\frac{p(x){\sum_{\tilde{x}}{p(y|\tilde{x})p^{(k)}{(\tilde{x})}}}}{p^{(k)}{(x)}{\sum_{\tilde{x}}{p(y|\tilde{x})p(\tilde{x})}}}]}\nonumber
\end{eqnarray}
est toujours positif et qu'il est nul si et seulement si\\
$p(x)=p^{(k)}{(x)}$
et $q(y)=q^{(k)}{(y)}$.\\
Le processus itératif devient alors:
\begin{small}
\label{our}
$$p^{(k+1)}{(x)} = \arg\max_{p(x)}\{I(p(x))-
\beta_k\{D(p(x)||p^{(k)}{(x)}) -D(q(y)||q^{(k)}{(y)})\}\}$$
\end{small}
A chaque itération, l'expression de $p^{(k+1)}{(x)}$ est la même que
dans (\ref{BA}). L'algorithme de Blahut-Arimoto s'interprète donc
comme un algorithme du point proximal dans lequel le paramètre
$\beta_k$ est constant et égal à 1.\\ L'approche intuitive de Matz
\cite{Matz04} consiste à remplacer la distribution de probabilité
$q(y)$ dans le terme de droite de l'équation précédente par la même
distribution $q^{(k)}{(y)}$ calculée à l'itération précédente.\\
Nous allons maintenant utiliser le degré de liberté supplémentaire
amené par $\beta_k$ pour augmenter la vitesse de convergence. Nous
choisissons $\beta_k$ comme suit:
$$\max_{\beta_k}{\beta_k{(D(p^{(k+1)}{(x)}||p^{(k)}{(x)})-D(q^{(k+1)}{(y)}||q^{(k)}{(y)}))}}$$
dans lequel $p^{(k+1)}{(x)}$ et $q^{(k+1)}{(y)}$ dépendent de
$\beta_k$. Cela guarantie que $I(p^{(k+1)}{(x)})-I(p^{(k)}{(x)})$
est maximale à chaque itération. Pour résoudre ce problème de
maximisation, nous avons utilisé la méthode de gradient conjugué qui
donne la valeur de $\beta_k$ la plus convenable en comparaison avec
l'approche proposée par Matz.
\subsubsection{Simulation} 
\begin{figure}
\begin{center}
\includegraphics[width= 0.3\textwidth,height=3.8cm]{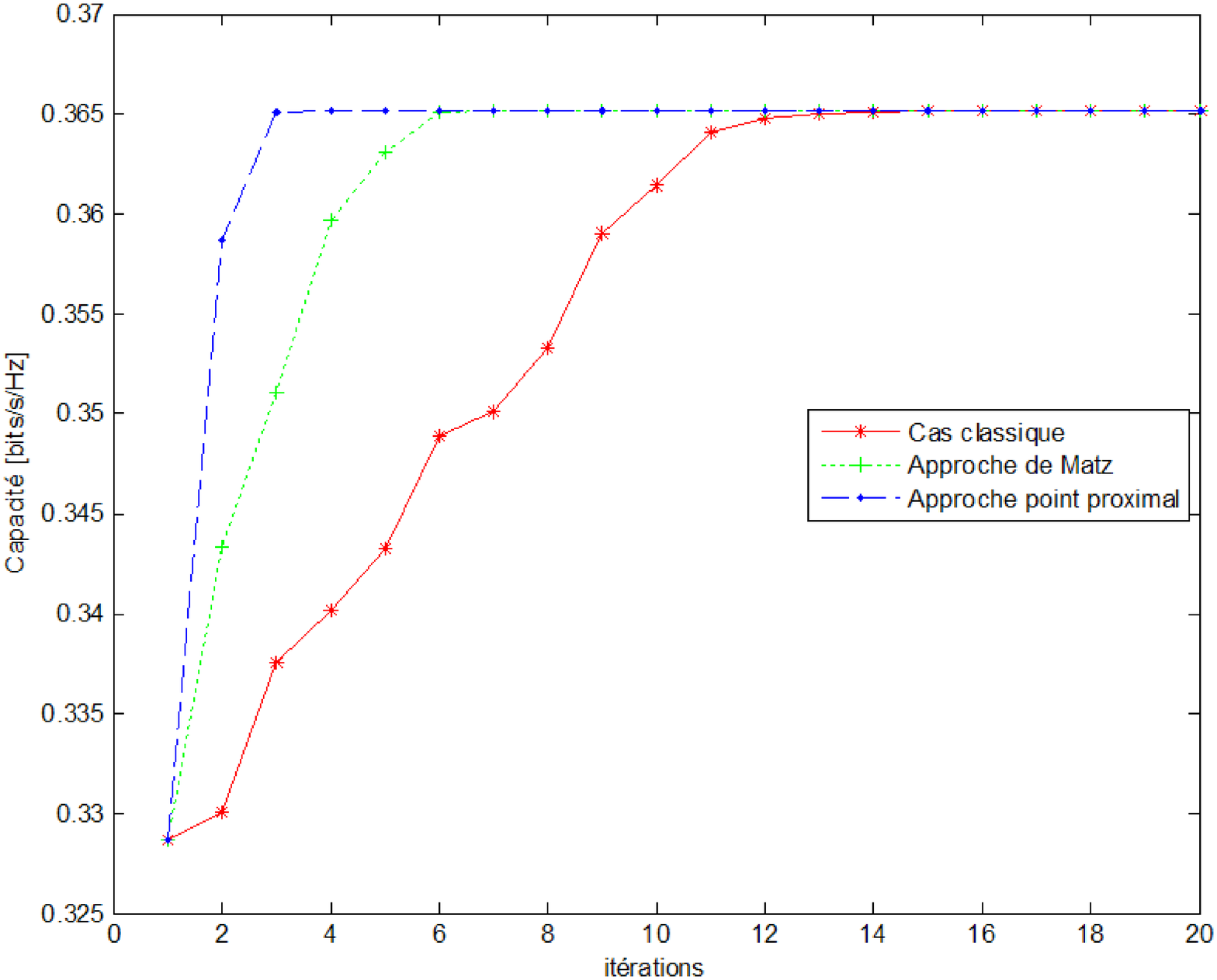}
\caption{\small{ Canal discret binaire symétrique.}}
\label{somefiglabel1}
\end{center}
\end{figure}

Nous testons les 3 algorithmes itératifs sur un canal discret binaire
symétrique défini par sa matrice de transition:
\begin{center}
\[Q= \left \{
\begin{array}{ccc}
   0.7 & 0.2 & 0.1 \\
   0.1 & 0.2 & 0.7
\end{array}
\right \}
\]
\end{center}
Les résultats (fig.\ref{somefiglabel1}) montrent que la capacité du
canal est atteinte après 20 itérations dans le cas classique, 7
itérations dans l'approche de Matz et 4 itérations dans notre cas
(avec une précision de $10^{-11}$).

\begin{figure}
\begin{center}
\includegraphics[width= 0.3\textwidth,height=3.8cm]{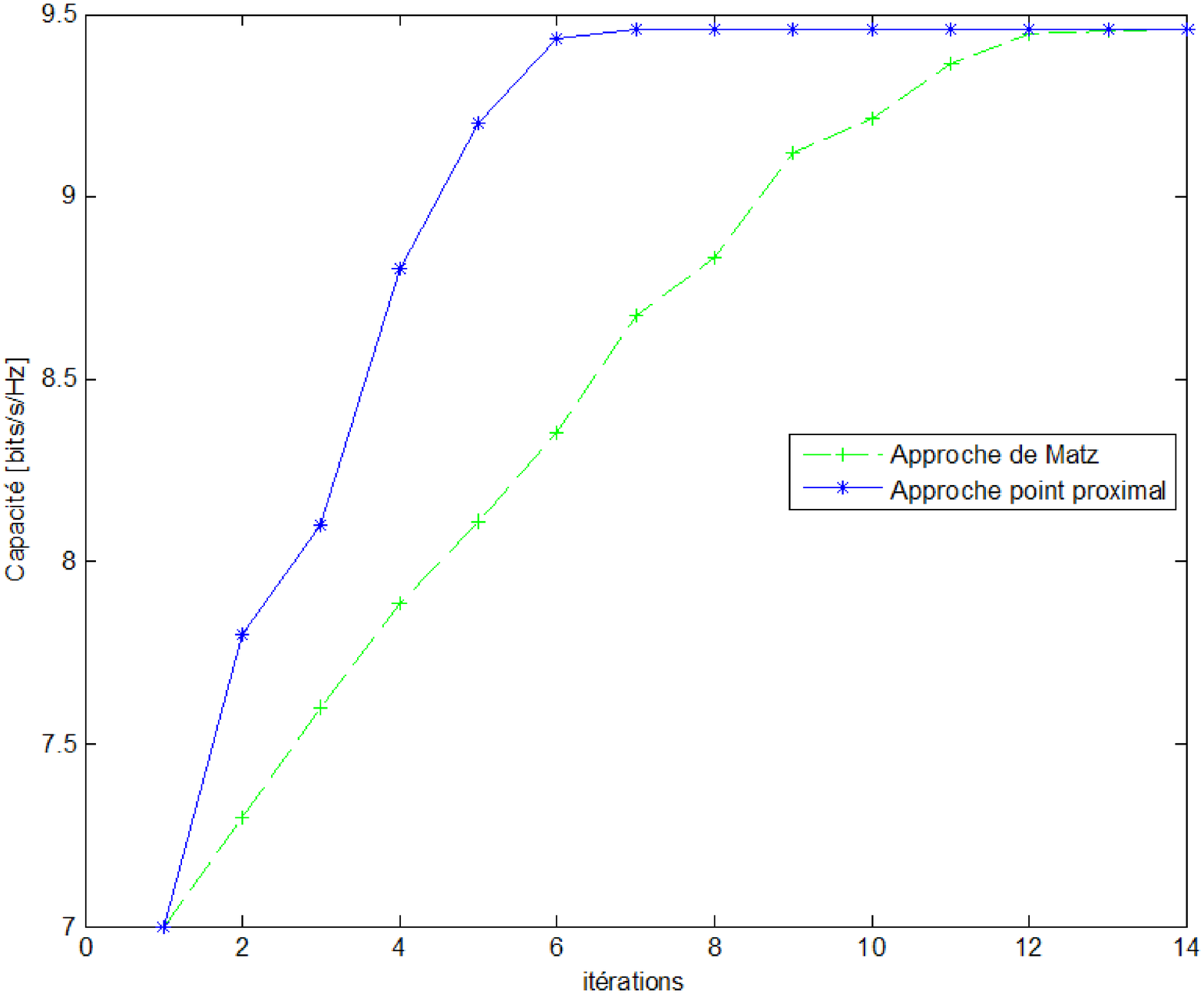}
\caption{\small{ Canal Gaussian Bernouilli-Gaussian
        ayant comme paramètres $(p=0.3,\sigma_b=0.01,
        \sigma_g=1)$}.}
        \label{somefiglabel}
\end{center}
\end{figure}

Nous comparons ensuite notre algorithme et celui de Matz dans le cas
d'un canal Gaussian Bernouilli-Gaussian dans le but de former une
matrice Q avec de grandes dimensions. Un tel canal est défini par :
$y_k=x_k+b_k+\gamma_k$
où
\begin{itemize}
\item
$b \sim \mathcal{N}(0,\sigma_b^2)$
\item
$\gamma_k=e_kg_k$ \hspace{0.5cm} avec \hspace{1cm} e : séquence de
Bernouilli(p)
\item
$g \sim \mathcal{N}(0,\sigma_g^2)$ \hspace{0.5cm} avec \hspace{1cm}
$\sigma_b^2\ll \sigma_g^2$
\end{itemize}
d'où \vspace{-0.4cm}
\begin{center}
$y_k=x_k+n_k$
\end{center}
avec \vspace{-0.4cm} \begin{center} $p(n_k)=(1-p)
\mathcal{N}(0,\sigma_b^2)+p \mathcal{N}(0,\sigma_b^2+\sigma_g^2)$
\end{center} La sortie $y_k$ a été discrétisée sur $40$ valeurs, et l'entrée $x_k$ sur $10$ valeurs.
Les résultats sont reportés sur la figure \ref{somefiglabel}. Nous
observons encore un gain conséquent grâce à notre approche.

\vspace*{-0.3cm}
\subsection{Outils de base}
Nous introduisons tout d'abord quelques notations. Soit ${\bf B}_i
\in \{0,1\}^N$ la représentation binaire d'un entier $i, 0\leq i\leq
2^{N-1}$. ${\bf B} = ({\bf B}_0,{\bf B}_1,...,{\bf B}_{2^N-1})^T$ de
dimension $2^N\times N$ est la matrice de la représentation binaire
de tous les mots de longueur N. Soit $\bf{\eta}$ la fonction densité
de probabilité de la variable $\chi={\bf B}_i$. On a donc
$$\bf{\eta}=(Pr[\chi={\bf B_0}],Pr[\chi=\bf{ B_1}],...,Pr[\chi=\bf{ B_{2^N-1}}])^T$$
Etant donné une fonction densité de probabilité $\eta$, ses
coordonnées logarithmiques sont le vecteur ${\bf \theta}$ dont le
$i^{eme}$ élément est donné par $\theta_i=\ln (Pr[\chi={\bf
B_{i}}])-\ln (Pr[\chi={\bf B_{0}}])$. Nous définissons aussi
$\lambda$ le vecteur des ratio dont l'élément j est défini par
$\lambda_j=log(\frac{Pr[\chi_j=1]}{Pr[\chi_j=0]})$ où $\chi_j$ est
le $j^{eme}$ bit du mot binaire $\chi$ et ${\bf \lambda}\in{\mathbb
R}^N$. Pour des densités séparables, c'est à dire qui sont égales au
produit des marginales, les coordonnées logarithmiques prennent la
forme ${\bf \theta}={\bf B}{\bf \lambda}$ \cite{Walsh}.
\subsubsection{Décodage itératif des modulations codées à bits
entrelacés {\cite{Caire}}}
%
\begin{figure}[!h]
\centerline{\epsfxsize=7cm\epsfysize=0.8cm\epsfbox{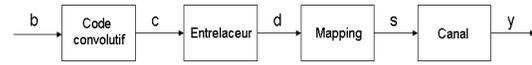}}
\caption{\footnotesize \label{result1} Codeur des modulations codées
à bits entrelacés.}
\end{figure}
%
\begin{figure}[!h]
\centerline{\epsfxsize=7cm\epsfysize=1.8cm\epsfbox{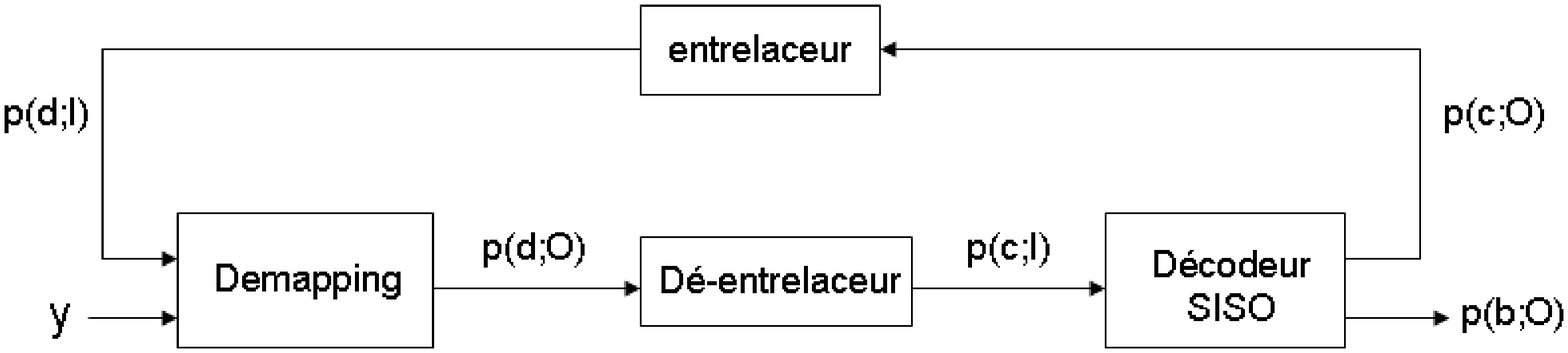}}
\caption{\footnotesize \label{result1} Décodeur itératif des
modulations codées à bits entrelacés.}
\end{figure}
Le décodage itératif pour les modulations codées à bits entrelacées
est constitué de deux blocs chacun ayant pour tâche d'évaluer des
probabilités a posteriori. Le premier bloc (demapping) contient les
informations concernant le mapping et le canal au travers de la loi
de probabilité $p({\bf y}\vert{\bf s})$ où ${\bf y}$ est le vecteur
reçu et ${\bf s}$ un vecteur de symbole. Ce bloc reçoit un a priori
(aussi appelé extrinsèque) qui lui est fourni par l'autre bloc. Il
est donc en mesure de fournir des probabilités à posterori que nous
noterons $p_{\bf B\lambda_1+\theta_m}$ où
$({\lambda_1})_{km+i}=\ln\left(\frac{p(d_{km+i}=1;I)}{p(d_{km+i}=0;I)}\right)$
est le vecteur contenant les log-ratio de la probabilité a priori
\cite{Walsh}. Le vecteur $\theta_m$ est le vecteur de coordonnées
logarithmiques obtenu à partir de $p({\bf y}\vert{\bf s})$. Le
second bloc contient les informations correspondant au codeur au
travers de la fonction indicatrice du code. Ce second bloc fournit
les probabilités a posteriori sur les bits $p_{\bf
B\lambda_2+\theta_c}$ où $\lambda_2$ dépend de l'a priori à l'entrée
du bloc et $\theta_c$ est le vecteur de coordonnées logarithmiques
obtenu à partir de la fonction indicatrice du code \cite{Walsh}. Par
ailleurs, l'a priori du bloc suivant est calculé en divisant la
probabilité a posteriori du bloc précédent par l'a priori qu'il a
reçu (propagation d'extrinsèques). Ce principe peut être résumé par
le processus itératif:
\begin{eqnarray}
\textnormal{Trouver} \; {\bf \lambda_2}^{(k+1)} \; \textnormal{telle que} \;\; p_{\bf B(\lambda_1^{(k)}+\lambda_2^{(k+1)})} = p_{\bf B\lambda_1^{(k)}+\theta_m} \label{eq mapp}\\
\textnormal{Trouver} \; {\bf \lambda_1}^{(k+1)} \; \textnormal{telle
que} \;\; p_{\bf B(\lambda_1^{(k+1)}+\lambda_2^{(k+1)})} = p_{\bf
B\lambda_2^{(k+1)}+\theta_c}\label{eq dec}
\end{eqnarray}
Ce processus itératif correspond à la résolution du problème de
minimisation suivant:\\Au niveau du demapping $$\min_{\bf
\lambda_2}D_{FD}(p_{\bf B\lambda_1+\theta_m},p_{\bf
B(\lambda_1+\lambda_2)})$$ Au niveau du décodeur
$$ \min_{\bf \lambda_1}D_{FD}(p_{\bf
B\lambda_2+\theta_c},p_{\bf B(\lambda_1+\lambda_2)})$$ Une solution
est satisfaisante si elle répond aux deux critères simultanément.\\
Cependant la minimisation de l'un de ces critères n'entraine pas
forcément la diminution de l'autre critère à
l'itération suivante. On peut donc craindre un comportement de l'algorithme. La méthode du point proximal 
permet de faire le lien entre les deux critères via le terme de
pénalité qu'elle introduit. Nous obtenons alors un nouveau processus
de minimisation:
\begin{eqnarray}
{\bf \lambda_2^{(k+1)}} =\min_{\bf \lambda_2} J_{\theta_m}({\bf \lambda_1},{\bf \lambda_2} )= \min_{\bf \lambda_2} D_{FD}({\bf p_{B\lambda_1+\theta_m}},{\bf p_{B(\lambda_1+\lambda_2)}})\nonumber\\
+\mu_m D_{FD}({\bf p_{B(\lambda_1^{(k)}+\lambda_2^{(k)})}},{\bf p_{B(\lambda_1+\lambda_2)}})\nonumber\\
{\bf \lambda_1^{(k+1)}} =\min_{\bf \lambda_1} J_{\theta_c}({\bf \lambda_1},{\bf \lambda_2} )=\min_{\bf \lambda_1} D_{FD}({\bf p_{B\lambda_2+\theta_c}},{\bf p_{B(\lambda_1+\lambda_2)}})\nonumber\\
+\mu_c D_{FD}({\bf p_{B(\lambda_1^{(k)}+\lambda_2^{(k+1)})}},{\bf
p_{B(\lambda_1+\lambda_2)}})\nonumber
\end{eqnarray}
Cela revient à trouver ${\bf \lambda_2^{(k+1)}}$ telle que
\begin{equation}
{\bf p_{B(\lambda_1^{(k)}+\lambda_2^{(k+1)})}} = \frac{{\bf
p_{B\lambda_1^{(k)}+\theta_m}}+\mu_m{\bf
p_{B(\lambda_1^{(k)}+\lambda_2^{(k)})}}}{1+\mu_m} \label{l2_pp}
\end{equation}
et  ${\bf \lambda_1^{(k+1)}} $ telle que
\begin{equation}{\bf p_{B(\lambda_1^{(k+1)}+\lambda_2^{(k+1)})}} = \frac{\bf p_{B\lambda_2^{(k+1)}+\theta_c}+\mu_c{\bf p_{B(\lambda_1^{(k)}+\lambda_2^{(k+1)})}}}{1+\mu_c}
\label{l1_pp}
\end{equation}
A la convergence, on retrouve les mêmes points stationnaires que
pour (\ref{eq mapp}) et (\ref{eq dec}). Pour assurer la décroissance
des fonctions de coût, nous choisissons $\mu_m$ et $\mu_c$ afin que\\
$J_{\theta_m}({\bf \lambda_1^{(k)}},{\bf \lambda_2^{(k+1)}} )\leq
J_{\theta_c}({\bf \lambda_1^{(k)}},{\bf \lambda_2^{(k)}} )$ et
\\$J_{\theta_c}({\bf \lambda_1^{(k+1)}},{\bf \lambda_2^{(k+1)}} )\leq
J_{\theta_m}({\bf \lambda_1^{(k+1)}},{\bf \lambda_2^{(k+1)}} )$.\\ La première inégalité est équivalente à\\
$J_{\theta_m}({\bf \lambda_1^{(k)}},{\bf \lambda_2^{(k+1)}} )\leq
\frac{\mu_m}{1+\mu_m}(D_{FD}({\bf
p_{B\lambda_1^{(k)}+\theta_m}},{\bf
p_{B(\lambda_1^{(k)}+\lambda_2^{(k)})}})+D_{FD}({\bf
p_{B(\lambda_1^{(k)}+\lambda_2^{(k)})}},{\bf
p_{B\lambda_1^{(k)}+\theta_m}}))$ car la distance de Fermi-Dirac est
convexe par rapport à son deuxième paramètre. D'autre part  $
D_{FD}({\bf p_{B\lambda_2^{(k)}+\theta_c}},{\bf
p_{B(\lambda_1^{(k)}+\lambda_2^{(k)})}})\leq J_{\theta_c}({\bf
\lambda_1^{(k)}},{\bf \lambda_2^{(k)}})$\\ D'après ces deux
relations, nous obtenons une borne supérieure pour $\mu_m$:
$$\mu_m\leq\frac{D_{FD}({\bf p_{B\lambda_2^{(k)}+\theta_c}},{\bf p_{B(\lambda_1^{(k)}+\lambda_2^{(k)})}})}{{\cal D}_{FD}-D_{FD}({\bf p_{B\lambda_2^{(k)}+\theta_c}},{\bf p_{B(\lambda_1^{(k)}+\lambda_2^{(k)})}})}$$
où ${\cal D}_{FD}$ est une distance symétrique:\\
${\cal D}_{FD}= D_{FD}({\bf p_{B\lambda_1^{(k)}+\theta_m}},{\bf
p_{B(\lambda_1^{(k)}+\lambda_2^{(k)})}})\\+D_{FD}({\bf
p_{B(\lambda_1^{(k)}+\lambda_2^{(k)})}},{\bf
p_{B\lambda_1^{(k)}+\theta_m}})$\\ La borne supérieure pour $\mu_c$
peut être obtenue d'une façon similaire. En itérant (\ref{l2_pp}) et
(\ref{l1_pp}) avec $\mu_c$ et $\mu_m$ choisis correctement nous
obtenons un algorithme qui converge vers les même points que le
décodage itératif classique (et qui a donc les mêmes performances en
terme de taux d'erreur binaire) tout en diminuant au fil des
itérations un critère désiré.
 \vspace*{-0.3cm}
\section{Conclusion}
Dans cet article, nous avons d'abord mis en évidence l'algorithme
itératif du point proximal. Nous avons ensuite présenté deux
algorithmes itératifs différents à la fois par l'application visée
et le processus itératif mis en jeu: l'algorithme itératif de
Blahut-Arimoto et l'algorithme de décodage itératif des modulations
codées à bits entrelacés. Une interprétation de ces deux algorithmes
basée sur la méthode de point proximal a donc été proposée appuyée
par des résultats de simulation. \vspace*{-0.3cm}
\bibliographystyle{IEEEtran}
\bibliography{Icassp09ziad}
\end{document}